\documentclass[10pt,conference]{IEEEtran}
\IEEEoverridecommandlockouts
\usepackage{cite}
\usepackage{amsmath,amssymb,amsfonts}
\usepackage{algorithmic}
\usepackage{balance}
\usepackage{graphicx}
\usepackage{textcomp}
\usepackage{tabularx}
\usepackage{xcolor}
\usepackage{multirow}
\usepackage{subfigure}
\def\BibTeX{{\rm B\kern-.05em{\sc i\kern-.025em b}\kern-.08em
    T\kern-.1667em\lower.7ex\hbox{E}\kern-.125emX}}

\title{Interpretable Battery Aging without Extra Tests via Neural-Assisted Physics-based Modelling\\
\thanks{This work was supported in part by 
A*STAR under its MTC Programmatic (Award M23L9b0052), MTC Individual Research Grants (IRG) (Award M23M6c0113), and MTC Programmatic (Award M24N6b0043).}
}

\makeatletter
\newcommand{\linebreakand}{%
  \end{@IEEEauthorhalign}
  \hfill\mbox{}\par
  \mbox{}\hfill\begin{@IEEEauthorhalign}
}
\makeatother

\author{
\IEEEauthorblockN{Yuan Qiu$^1$, Wei Li$^1$, Wei Zhang$^{1,*}$\thanks{$^*$ Corresponding author. Email: wei.zhang@singaporetech.edu.sg.}, Yi Zhou$^1$, Fang Liu$^2$, Jianbiao Wang$^3$, Zhi Wei Seh$^3$}
\IEEEauthorblockA{
\textit{$^1$Singapore Institute of Technology (SIT), Singapore, 828608}
\\\textit{$^2$Singapore University of Social Sciences (SUSS), Singapore, 599494}
\\\textit{$^3$Institute of Materials Research and Engineering (IMRE), Agency for Science, Technology and Research (A*STAR),}
\\\textit{2 Fusionopolis Way, Innovis \#08-03, Singapore 138634, Republic of Singapore}
\\\{yuan.qiu, wei.li, wei.zhang, zhouyi\}@singaporetech.edu.sg, liufang@suss.edu.sg, \{wang\_jianbiao, sehzw\}@a-star.edu.sg}
}

\begin{document}
\bstctlcite{IEEEexample:BSTcontrol}

\maketitle

\begin{abstract}
State of health (SoH) is widely used for battery management, but it is a single scalar and offers limited interpretability. Two batteries with similar SoH can exhibit very different degradation behaviors and the lack of interpretability hinders optimal battery operation. In this paper, we propose \textsc{IBAM} for \underline{i}nterpretable \underline{b}attery \underline{a}ging \underline{m}odelling with a neural-assisted physics-based framework. \textsc{IBAM} outputs a 2-D aging fingerprint without extra diagnostic tests and uses only routine logs from the battery management system. The fingerprint offers great interpretability by capturing a battery's curve-wide polarization voltage loss and the tail loss near the end-of-discharge. \textsc{IBAM} first creates a physics-based battery model based on a fractional-order equivalent-circuit model, and then extracts per-cycle fingerprints from the model using a two-stage least-squares method. \textsc{IBAM} further anchors fingerprints on the SoH axis with physics-guided regression, where the per-cycle SoH is estimated via a bidirectional gated recurrent unit with customized multi-channel voltage features. Across batteries with short-, medium-, and long-lifespans, \textsc{IBAM} consistently yields the best physics-model fidelity at different aging stages, and provides clear interpretations of degradation mechanisms and fingerprint patterns about batteries of different lifespans. The resulting fingerprints support interpretable battery health assessment and can inform battery control choices.
\end{abstract}

\begin{IEEEkeywords}
Battery health, state of health, machine learning, industrial artificial intelligence, interpretability, explainability 
\end{IEEEkeywords}

\section{Introduction}
Electric vehicles (EVs) are becoming a mainstream transport option. The International Energy Agency (IEA) reports that EV sales exceeded 17 million globally in 2024, reaching a sales share of over 20\% of new cars worldwide \cite{IEA25}. The battery is one of the most important and expensive components in an EV. Its performance and degradation therefore become practical concerns, as battery aging directly affects range, power capability, charging limits, etc. Battery operation is managed by the battery management system (BMS), where health monitoring is foundational. A widely used health indicator is the state of health (SoH), which is commonly defined as the ratio of the current usable capacity to the rated beginning-of-life capacity. However, SoH is not directly observable during operation and remains challenging to estimate and forecast accurately and robustly under diverse battery usage conditions.

Most battery health studies focus primarily on the accuracy of SoH modelling and prediction. A common trend is to adopt data-driven machine learning (ML) algorithms, with recent efforts increasingly favor large and complex ML models to maximize accuracy. Basic ML models are often considered in early studies, e.g., regression in \cite{severson2019data} and ensemble learning in \cite{kokkalis2024predicting}. Hybrid ML architectures combine complementary backbones, e.g., Transformer and bidirectional gated recurrent unit (BiGRU) in \cite{liu2025data} and Informer and gated recurrent unit in \cite{sameer2025ginet}. Recent advancements mainly follow two notable directions. One direction integrates domain knowledge and physics to improve generalization and reduce reliance on purely ML-based modelling, e.g., guiding liquid neural networks with thermal-related entropy in \cite{li2026entrolnn} and expanding temporal convolutional networks with the features from the equivalent circuit model (ECM) in \cite{sameer2025pace}. The other direction explores generative AI (GenAI) for battery health modelling, e.g., \cite{eivazi2024diffbatt} uses diffusion modelling for estimating remaining useful life and \cite{kwan5871665knowdiff} regulates GenAI with battery knowledge for robust SoH prediction. While these approaches can be highly accurate, they typically output a single health scalar and provide limited visibility into degradation mechanisms and the evolution of these mechanisms. Such a lack of explainability and interpretability make it difficult to translate a prediction of battery health into actionable battery management decisions, and motivates interpretable battery analytics.

A small but growing line of work argues that accuracy alone is insufficient and battery health models should also provide explanations that battery operators and users can trust and act on. Most existing studies follow a post-hoc explainable AI paradigm on ML-based models. A unified pipeline that combines post-hoc explainability with available tools such as Shapley additive explanations (SHAP) and local interpretable model-agnostic explanations is proposed in \cite{lim2025unified} for feature-level interpretation and confidence-aware diagnostics. Several recent studies also extend SHAP-based interpretation from explaining a single health prediction to summarizing degradation behavior over time. For example, SHAP is aggregated across sliding-window sequences in \cite{gahramanov2025detecting} to enable identification of dominant features, and used to interpret multiple ML models in \cite{tetik2026feature}. Another group of studies improves explainability by introducing additional sensing modalities that are directly coupled with batteries' internal electrochemical and mechanical changes. For example, a study extracts ultrasound features as inputs for SoH models and then uses interpretable models and explainability AI tools to visualize the relationship between the sensing features and battery degradation \cite{liu2024ultrasonic}. Overall, current explainable battery studies mainly improve transparency by attaching explanation tools to ML models or by adding extra sensing modalities. The proposed explanations are also not explicitly physics-guided so the explanations may not reflect real degradation behavior.

In this paper, we aim to enable interpretable battery health analytics from routine BMS logs without introducing additional battery diagnostic tests. We propose \textsc{IBAM}, an \underline{i}nterpretable \underline{b}attery \underline{a}ging \underline{m}onitoring framework. \textsc{IBAM} models battery operation dynamics with a fractional-order ECM (FOECM), which is physics-based and encodes battery knowledge in a compact set of interpretable parameters. The physics model includes two resistance-like parameters $R_{\mathrm{dyn}}$ and $R_{\mathrm{W}}$. The former captures the broad under-load voltage penalty, where the discharge voltage curve of a battery shifts downward and drops faster as the battery ages. The latter captures the late-stage voltage drop that emerges near the end-of-discharge and can trigger earlier low-voltage cutoff. These two parameters summarize complementary degradation effects in a compact and physically meaningful way, so \textsc{IBAM} defines $(R_{\mathrm{dyn}}, R_{\mathrm{W}})$ as an aging fingerprint for interpreting a battery's health beyond the commonly used SoH. \textsc{IBAM} first fits the model to each discharge cycle using a two-stage least squares method (LSM) to extract per-cycle raw fingerprints associated with two voltage-loss mechanisms. \textsc{IBAM} then estimates cycle-wise SoH using a BiGRU model with multi-channel features extracted from the same discharge voltage trace. Finally, \textsc{IBAM} performs a physics-guided mapping that projects the cycle-wise fingerprints onto the SoH domain via isotonic regression, so that the fingerprints are anchored to the industry-standard health axis. This yields physically consistent SoH-indexed fingerprints that can be queried online to support BMS decision-making. For example, a high $R_{\mathrm{dyn}}$ suggests dominant polarization loss and motivates conservative high-rate discharging, while a high $R_{\mathrm{W}}$ indicates pronounced end-of-discharge voltage loss and suggests stricter voltage gating policies. In summary, we have the following main contributions in this paper.
\begin{itemize}
    \item We formulate interpretable battery health monitoring from routine BMS logs, and interpret SoH using a 2-D aging fingerprint that captures polarization and tail losses.
    \item We propose \textsc{IBAM}, a physics-based framework that reliably identifies this fingerprint and constructs physically consistent SoH-indexed fingerprints for online querying.
    \item We validate \textsc{IBAM} on short-, medium-, and long-life batteries. \textsc{IBAM} achieves consistently improved physics-model fidelity and reveals lifespan-aligned fingerprint patterns that yield actionable degradation insights.
\end{itemize}

The rest of the paper is organized as follows. We present the technical details of \textsc{IBAM} in Section \ref{sec:sys}. In Section \ref{sec:exp}, we conduct an experimental study and present the experiment results. Finally, we conclude this paper in Section \ref{sec:conclusion}.

\section{\textsc{IBAM} Methodology}
\label{sec:sys}
\begin{figure}
    \centering	\includegraphics[width=0.98\linewidth]{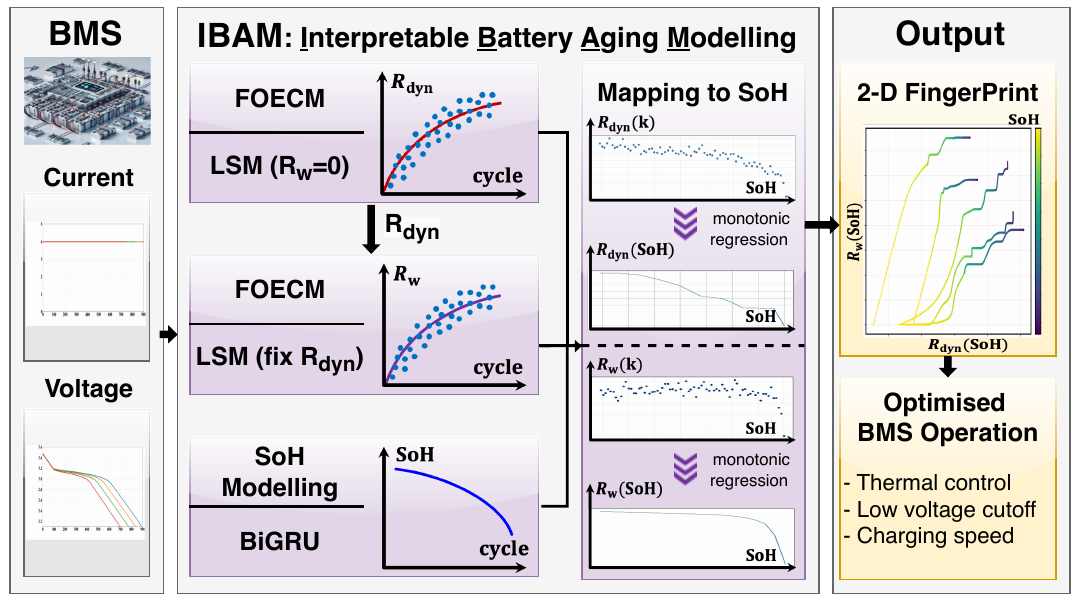}
    \caption{An illustration of \textsc{IBAM} architecture.}
    \label{fig:sys}
\end{figure}

An overview of \textsc{IBAM}'s system architecture is in Fig. \ref{fig:sys}, and technical details are provided in the following subsections. 

\subsection{Problem Formulation and Interpretation Targets}
This paper adopts a \emph{no-extra-tests} setting. We use only the signals and sensor measurements that are routinely logged by a typical BMS during battery operation. We do not assume additional cycling procedures, diagnostic tests, or sensor installation for practical deployments. We consider a battery observed over $n_c$ charge/discharge cycles, and the BMS log for cycle $c$ is given by,
\begin{equation}
\mathcal{D}_c=\Big\{\big(t_i,\mathbf{x}_c(t_i)\big)\Big\}_{i=1}^{T_c},
\end{equation}
where $t_i$ denotes the $i$-th timestamp within cycle $c$ and $T_c$ is the number of timestamps in the cycle. The log of each timestamp is recorded as a 2-tuple, e.g., $\big(t_i,\mathbf{x}_c(t_i)\big)$, where $\mathbf{x}_c(t)$ stacks the BMS-available channels such as terminal voltage, current, and temperature when available. We assume that the logged channels are time-aligned by the BMS, e.g., signals can be resampled or interpolated onto a common time grid. In this study, we primarily rely on voltage trajectory $V_c(t)$, which carries the dominant information about cycle-to-cycle degradation. Other BMS channels in $\mathbf{x}_c(t)$ such as current and temperature are used when available for standard state calculation and calibration. For example, current $I_c(t)$ is needed for state of charge (SoC) computation. It is approximately constant in many practical settings, and it contributes limited additional variability compared to the voltage trace.

In this paper, we have two complementary goals for battery health monitoring. Our first goal is to quantify the battery's health for each cycle, and the second goal is to interpret that health estimate by providing physically meaningful interpretations of the degradation effects. First, we follow common practice and use SoH as the primary cycle-level indicator of degradation. SoH measures the remaining usable capacity relative to a fresh cell and is typically defined as $Q_t /Q_0 \times 100\%$ as a percentage for the SoH at time $t$, where $Q_t$ denotes the usable capacity at $t$ and $Q_0$ is the rated capacity at the beginning of life. While SoH provides a concise scalar summary of degradation, it does not reveal degradation mechanisms underlying a specific health level. As such, our second goal, also the main goal of this paper, is to go beyond SoH and interpret it using factors with consistent physical meaning. 

\begin{figure}[t]
    \centering	\includegraphics[width=0.9\linewidth]{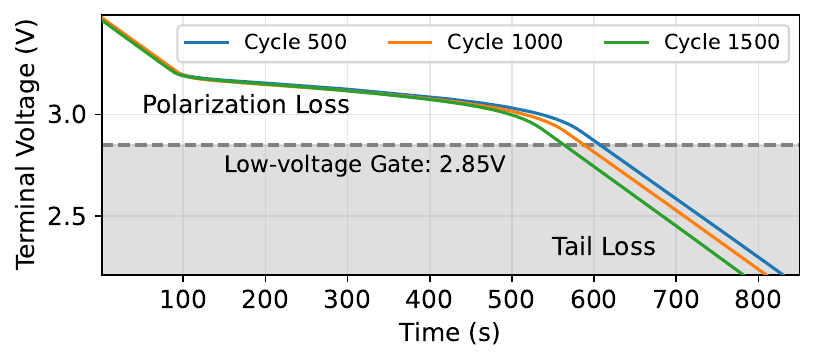}
    \caption{Illustration of the polarization loss and tail loss for a representative battery cell across different discharge cycles.}
    \label{fig:loss}
\end{figure}
\begin{figure}
    \centering	
    \def \tmph{1.05in}
    \subfigure[ECM]{\includegraphics[height=\tmph]{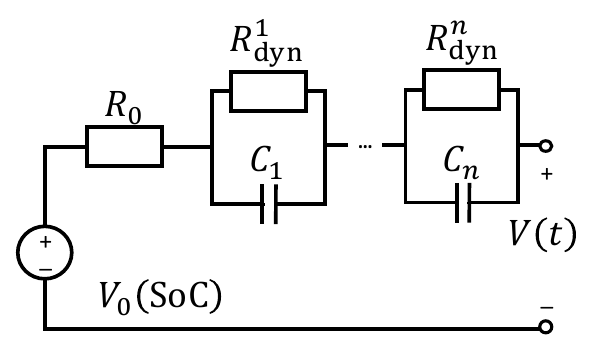}\label{fig:ecm1}}
    \hspace{-0.5em}
    \subfigure[FOECM]{\includegraphics[height=\tmph]{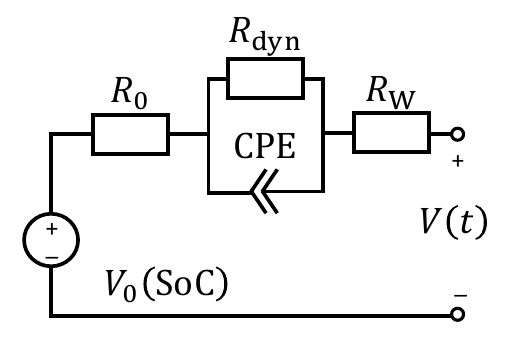}\label{fig:ecm2}}
    \caption{Illustration of the physics-based models for ECM and FOECM.}
    \label{fig:ecm}
\end{figure}

In practice, two batteries can share a similar SoH yet behave quite differently in operation. Motivated by common observations in battery health studies, degradation shown in routine discharge voltage curves typically appears in two distinct ways. One is the curve-wide under-load loss, where battery aging often presents as a global downward shift across discharge curves from different cycles. This indicates increasing under-load voltage loss and decreasing power capability, often described as polarization loss, as shown in Fig. \ref{fig:loss}. As this loss grows, the battery becomes more prone to voltage drop and heat generation under high current, since resistive heating scales with current squared. The other is the tail loss. Aging can be more pronounced near the end-of-discharge per cycle, where the voltage drops below the minimum required voltage and approaches the low-voltage cutoff earlier. Part of the remaining energy becomes infeasible to use under practical voltage constraints. As such, a single SoH value can appear optimistic even though end-of-discharge usability deteriorates. 

Accordingly, we represent battery health as a compact 2-D aging fingerprint $\mathbf{z} = (z^1, z^2)$, where $z^1$ and $z^2$ quantify the polarization loss and tail loss, respectively. We use two views of this fingerprint. In the cycle domain, we form a discrete sequence $\{(z^1_c, z^2_c)\}_{c=1}^{n_c}$ as the fingerprints across different cycles. In the SoH domain, we re-parameterize the fingerprints as health-conditioned curves $\mathbf{z}(s)=\big(z^1(s), z^2(s)\big)$ with SoH $s$ as the independent variable. Fingerprints in this domain enable noise-robust interpretations across the aging trajectory. 

\subsection{Physics-based Model for Fingerprint Extraction}
To extract an interpretable aging fingerprint from routine BMS logs, we adopt a physics-based battery model. Such a model provides a structured decomposition of terminal voltage into components with clear physical meaning, which directly supports our goal of quantifying the two voltage-related degradation mechanisms. Moreover, because the model structure is grounded in established battery domain knowledge, its key parameters can be identified from standard operating data without requiring extensive historical data or additional diagnostic experiments. Specifically, we introduce ECM and its fractional-order extension in this study. ECM first serves as a physics-grounded baseline for describing the discharge voltage under load \cite{sameer2025pace}. It represents the terminal voltage as an OCV component minus an ohmic drop and a set of polarization or relaxation drops, as shown in Fig. \ref{fig:ecm1}, as,
\begin{equation}
V(t) = V_0\big(\mathrm{SoC}(t)\big) - I(t)R_0 - \sum\nolimits_{i=1}^{m} v_i(t),
\label{eq:ecm}
\end{equation}
where $V_0(\cdot)$ is the open-circuit voltage (OCV), which changes as the battery is discharged and depends on SoC. SoC can be computed by standard bookkeeping, e.g., $\mathrm{SoC}(t) = \mathrm{SoC}(t_0) - \frac{1}{Q}\int_{t_0}^{t} I(\tau) \mathrm{d}\tau$ where  $t_0$ is the start time and $Q$ is the battery's usable capacity for SoC normalization. In addition, $R_0$ is the ohmic resistance, often modeled as a constant \cite{xie2025impedance}. Each $v_i(t)$ corresponds to the voltage drop across the $i$-th resistor–capacitor pair. For each pair, the dynamics follow the standard relation as,
\begin{equation}
\frac{\mathrm{d} v_i(t)}{\mathrm{d}t} = -\frac{1}{R_i C_i}v_i(t) + \frac{1}{C_i} I(t),
\label{eq:ecm_rc}
\end{equation}
where $I(t)$ is the applied current at time $t$, and $R_i$ and $C_i$ control how quickly the pair voltage responds and relaxes under current changes. The multiple resistor–capacitor pairs capture relaxation dynamics under the applied current $I(t)$, and their aggregate contribution, as shown in the last term of Eq. (\ref{eq:ecm}), forms a load-dependent voltage loss that shapes the terminal voltage over a broad portion of the discharge trajectory. While ECM is partially interpretable and widely used, representing multi-timescale relaxation such as tail loss may require many resistor–capacitor pairs and can still be insufficient for capturing aging behaviors. This motivates FOECM, which provides a compact yet flexible description of both the curve-wide polarization loss and the end-of-discharge tail loss. Compared with Eq. (\ref{eq:ecm}), FOECM preserves the same high-level voltage decomposition, i.e., OCV minus under-load losses, but refines the polarization-related voltage drop and introduces an explicit term for the tail loss. We therefore rewrite the terminal-voltage model as shown in Fig. \ref{fig:ecm2} as,
\begin{equation}
V(t) = V_0\big(\mathrm{SoC}(t)\big) - I(t)R_0 - v_{\mathrm{dyn}}(t) - v_{\mathrm{W}}(t),
\label{eq:foecm}
\end{equation}
where $v_{\mathrm{dyn}}(t)$ denotes the polarization-related loss term and $v_{\mathrm{W}}(t)$ denotes the tail-loss term. In FOECM, the term $v_{\mathrm{dyn}}(t)$ is captured by a polarization pair composed of a resistor $R_{\mathrm{dyn}}$ in parallel with a constant phase element (CPE). $v_{\mathrm{dyn}}(t)$ is defined as the voltage across the parallel resistor and CPE paths and we denote the currents through the two paths by $I_{\mathrm{R}}(t)$ and $I_{\mathrm{CPE}}(t)$, for the resistor and CPE, respectively. By Kirchhoff's laws, we have,
\begin{equation}
I(t) = I_{\mathrm{R}}(t) + I_{\mathrm{CPE}}(t).
\end{equation}
Since the resistor and CPE are in parallel, they share the same voltage $v_{\mathrm{dyn}}(t)$. Based on Ohm's law, we have,
\begin{equation}
I(t) = v_{\mathrm{dyn}}(t) /R_{\mathrm{dyn}} + I_{\mathrm{CPE}}(t),
\end{equation}
which makes the role of $R_{\mathrm{dyn}}$ explicit. For a given applied current $I(t)$, a large $R_{\mathrm{dyn}}$ requires a large voltage to sustain the resistor path current, and increases the polarization-based voltage drop for the terminal voltage. We therefore use $R_{\mathrm{dyn}}$ as the FOECM parameter that quantifies the polarization loss of the aging fingerprint. Note that CPE captures a phase-lag response and therefore does not react instantaneously to voltage changes. At this stage, it suffices to note that $I_{\mathrm{CPE}}(t)$ is also driven by the same voltage $v_{\mathrm{dyn}}(t)$ and our formulation does not depend on the CPE's exact form. 

In FOECM, we model the tail loss term with a Warburg-type diffusion element to capture the fact that, near the end-of-discharge, charge carriers such as lithium ions cannot transport fast enough to sustain the load, which causes an extra voltage drop $v_{\mathrm{W}}(t)$. Because the element lies on the main current path, the current through the tail element equals the terminal current, i.e., $I_{\mathrm{W}}(t) = I(t)$. Here, we parameterize the tail element with a resistance-like magnitude $R_{\mathrm{W}}$ and a diffusion timescale $\tau_{\mathrm{W}}$. Although \textsc{IBAM} mainly uses time-domain battery trajectories, it is convenient to specify the Warburg element using its standard frequency-domain impedance form as,
\begin{equation}
Z_{\mathrm{W}}(j\omega) = R_{\mathrm{W}} / \sqrt{j\omega \tau_{\mathrm{W}}},
\label{eq:warburg_norm}
\end{equation}
where $Z_{\mathrm{W}}(j\omega)$ is the Warburg impedance, $j=\sqrt{-1}$ is the imaginary unit, and $\omega$ is the angular frequency. Here $Z_{\mathrm{W}}(j\omega)$ specifies the frequency-domain impedance of the tail element, and describes the relationship between a current component at frequency $\omega$ and a corresponding voltage drop across the element. Equivalently, in the time domain, this means the tail voltage $v_{\mathrm{W}}(t)$ depends on the current history, rather than only the instantaneous current. The equation highlights the distinct roles of $R_{\mathrm{W}}$ and $\tau_{\mathrm{W}}$, where the former directly scales the magnitude of the voltage drop $v_{\mathrm{W}}(t)$ under a given current profile and the latter sets the diffusion timescale, e.g., slow diffusion with large $\tau_{\mathrm{W}}$. Consequently, $R_{\mathrm{W}}$ scales with the tail loss  and we therefore use it to quantify the loss. In summary, FOECM instantiates the two voltage-based battery degradation effects with two role-specific parameters $R_{\mathrm{dyn}}$ and $R_{\mathrm{W}}$. The parameters govern the curve-wide polarization loss and end-of-discharge tail loss, respectively, and we therefore instantiate $(z^1,z^2)$ in the 2-D fingerprint with the FOECM's physically-interpretable $(R_{\mathrm{dyn}},R_{\mathrm{W}})$.

\subsection{Per-cycle Fingerprint Identification with Two-Stage LSM}
We identify the cycle-wise FOECM fingerprint parameters from routine BMS logs on a per-cycle basis. In this study, the aging fingerprint is instantiated by two FOECM parameters $R_{\mathrm{dyn}}$ and $R_{\mathrm{W}}$, and the rest of the FOECM parameters such as $V_0(\cdot)$, $\mathrm{SoC}(t)$, and $R_0$ are obtained from standard calibration or treated as fixed inputs. Given a discharge cycle $c$ with measured voltage and current $\{V_c(t_i), I_c(t_i)\}_{i=1}^{T_c}$, our goal is to find parameter values that make the FOECM-predicted voltage best match the measured voltage. We achieve this by LSM, which plays the role of physics-based parameter fitting. Because the polarization loss and tail loss affect different regions of the voltage trajectory, we adopt a two-stage strategy to separate the contributions of the two loss terms. In the first stage, we fit the polarization term over the entire discharge curve where the tail term is temporarily disabled. In the next stage, we fit the tail loss using the remaining voltage residual after subtracting the stage-1 predicted polarization voltage, so as to focus on the end-of-discharge behavior. 

In the first stage, we estimated $R_{\mathrm{dyn}}$ by fitting the polarization voltage drop assuming $v_{\mathrm{W}}(t)=0$. For simplicity, we treat FOECM as a cycle-level forward model $\mathcal{F}(\cdot)$, which takes the measured current trajectory of a discharge cycle and produces a predicted terminal voltage trajectory. Specifically, we have,
\begin{equation}
\hat{\mathbf{V}}^1_c(R_{\mathrm{dyn}})\triangleq \mathcal{F}(\mathbf{I}_c;R_{\mathrm{dyn}},\Theta^1),
\end{equation}
where the model specification includes the following components. We use the measured current time series $\mathbf{I}_c=\{I_c(t_i)\}_{i=1}^{T_c}$ of cycle $c$. $R_{\mathrm{dyn}}$ is the only cycle-wise parameter estimated at this stage. $\Theta^1$ collects all fixed or known settings such as $V_0(\cdot)$ and $R_0$ shared across cycles. Any internal dynamic states of FOECM such as internal polarization/CPE states are initialized at the start of each cycle and then calculated forward under the input current. The model output is the voltage curve $\hat{\mathbf{V}}^1_c(R_{\mathrm{dyn}})$ predicted by FOECM at the first stage when the polarization loss is parameterized by $R_{\mathrm{dyn}}$. Essentially, we aim to apply LSM to estimate the cycle-wise fingerprint parameter $R_{\mathrm{dyn}}$ by comparing the model output voltage $\hat{\mathbf{V}}^1_c(R_{\mathrm{dyn}})$ against the measured voltage $\mathbf{V}_c=\{V_c(t_i)\}_{i=1}^{T_c}$. The optimal $R_{\mathrm{dyn}}^c$ can be derived by,
\begin{equation}
R_{\mathrm{dyn}}^c=\arg\min_{R_{\mathrm{dyn}}}
\left\|\mathbf{V}_c-\hat{\mathbf{V}}^1_c(R_{\mathrm{dyn}})\right\|_2^2,
\end{equation}
where $\|\cdot\|_2^2$ is the sum of squared errors and the optimization is equivalent to minimizing the mean squared error.

In the second stage, we estimate $R_{\mathrm{W}}$ by fitting the tail loss term. We utilize the same BMS logs for current $\mathbf{I}_c$ and voltage $\mathbf{V}_c$ as the first stage. As tail loss is most pronounced near the end-of-discharge, we define a low-voltage gate to select tail-region timestamps for cycle $c$ as, 
\begin{equation}
\mathcal{T}_c \triangleq \{i \mid 1 \leq i \leq T_c \land V_c(t_i)\le V_{\mathrm{g}}\},
\end{equation}
where $V_{\mathrm{g}}$ is a fixed low-voltage threshold. We use $\mathbf{I}_c[\mathcal{T}_c]$ and $\mathbf{V}_c[\mathcal{T}_c]$ to denote the sub-vectors restricted to these gated timestamps for current and voltage, respectively. With the polarization parameter fixed to the optimal $R_{\mathrm{dyn}}^c$ estimated in the first stage, we activate the tail term in FOECM by removing the constraint of $v_{\mathrm{W}}(t)=0$ and define an updated voltage predictor as,
\begin{equation}
\hat{\mathbf{V}}^2_c(R_{\mathrm{W}})\triangleq \mathcal{F}(\mathbf{I}_c;R_{\mathrm{dyn}}^c,R_{\mathrm{W}},\Theta^2),
\end{equation}
where $R_{\mathrm{W}}$ is the only cycle-wise parameter estimated at this stage. $\Theta^2$ contains the same fixed or known terms as $\Theta^1$ with minor additions to support the tail-region fitting. We then apply LSM again by comparing the measured and predicted terminal voltage curves on the gated timestamps as,
\begin{equation}
R_{\mathrm{W}}^c=\arg\min_{R_{\mathrm{W}}}
\left\|\mathbf{V}_c[\mathcal{T}_c]-\hat{\mathbf{V}}^2_c(R_{\mathrm{W}})[\mathcal{T}_c]\right\|_2^2,
\end{equation}
which yields the cycle-wise estimated $R_{\mathrm{W}}^c$ for cycle $c$. Altogether, we have the cycle $c$ fingerprint $(R_{\mathrm{dyn}}^c,R_{\mathrm{W}}^c)$ to reflect the battery's degradation mechanisms for cycle $c$. As the battery continues operating, we apply the same per-cycle fingerprint identification online and maintain the fingerprint history up to the current cycle, i.e., $\{(R_{\mathrm{dyn}}^i,R_{\mathrm{W}}^i)\}_{i=1}^c$. 

\subsection{Physics-guided SoH-domain Fingerprint Mapping}
While our two-stage LSM provides cycle-wise aging fingerprints, these estimations can fluctuate from cycle to cycle due to measurement noise and operating variability. Moreover, the cycle index itself is not a industry-standard battery health scale. To obtain a more interpretable and transferable representation, we map the fingerprints from the cycle domain to the SoH domain as $\big(R_{\mathrm{dyn}}(\mathrm{s}),R_{\mathrm{W}}(\mathrm{s})\big)$ for different SoH $s\leq 100\%$. Because SoH is not directly available in BMS logs, we predict it from the discharge voltage using a BiGRU-based estimator. With the predicted SoH, we then learn a physics-guided mapping that converts the raw cycle-wise fingerprints into physics-guided and interpretable SoH-domain curves.

\subsubsection{BiGRU-based SoH Modelling}
To enable SoH-domain fingerprinting, we first estimate a battery's SoH from routine discharge voltage data. We adopt a lightweight sequence regressor based on a two-layer BiGRU model which maps each cycle's voltage curve to a scalar SoH. Specifically, we have voltage $\{V_c(t_i)\}_{i=1}^{T_c}$ for cycle $c$ and we construct a 4-channel time-series input. Since different cycles may have different sampling rates and lengths, we resample the raw trajectories onto a fixed-length time grid and construct the 4-channel input on this standardized grid via linear interpolation. The channels include the terminal voltage $V_c(t)$, the relative terminal voltage to OCV, the voltage change rate, and the timestamps. Each BiGRU layer uses 96 hidden units per direction with dropout and the output at the final timestamp is used as a cycle-level representation. A lightweight regression head with two linear layers and a \texttt{ReLU} activation in between maps the representation to a latent value which is processed by a \texttt{Sigmoid} function to generate a valid SoH value. During battery operation, once cycle $c$ is observed, the BiGRU model outputs the predicted SoH $\hat{s}_c$, which serves as the health coordinate for the subsequent SoH-domain fingerprint mapping. Note that BiGRU is used as a practical SoH model for \textsc{IBAM} as it offers a good balance between lightweight deployment and competitive modelling capability. Our team's research portfolio includes more sophisticated ML-based SoH modelling architectures, e.g., \cite{kwan5871665knowdiff,sameer2025pace,li2026entrolnn}, which, however, are not adopted to keep the overall \textsc{IBAM} pipeline efficient.

\subsubsection{Physics-guided Mapping}
At the current cycle $c$, we have accumulated cycle-wise SoH estimates $\{\hat{s}_i\}_{i=1}^{c}$ from the BiGRU model and cycle-wise raw fingerprint estimates $\{R_{\mathrm{dyn}}^i\}_{i=1}^{c}$ and $\{R_{\mathrm{W}}^i\}_{i=1}^{c}$ from the two-stage LSM. Because the cycle index is not a comparable health scale across batteries, we convert these cycle-wise points into SoH-domain mappings. We form paired samples $(\hat{s}_i,R_{\mathrm{dyn}}^i)$ and $(\hat{s}_i,R_{\mathrm{W}}^i)$ for cycle $1\leq i\leq c$. For intuitive understanding, these pairs form $c$ points in a two-dimensional coordinate system, e.g., SoH as the horizontal axis and $R_{\mathrm{dyn}}/R_{\mathrm{W}}$ as the vertical axis. 

Our first task is to construct an SoH-domain mapping $R_{\mathrm{dyn}}(s)$ that follows a physically consistent monotonic trajectory and stays close to the obtained points. Starting from the paired samples $\{(\hat{s}_i, R_{\mathrm{dyn}}^{i})\}_{i=1}^{c}$, we do not treat all points equally because the per-cycle identification of $R_{\mathrm{dyn}}^{i}$ can be affected by measurement noise and operating variability. We therefore associate each cycle $i$ with a scalar error computed from the per-cycle FOECM voltage fit, and assign a large weight to a point when this error is small. Let $\varepsilon_i \ge 0$ denote the per-cycle squared error for cycle $i$, and we set the weight as $w_i= 1 / (\varepsilon_i+\varepsilon_0)$, where $\varepsilon_0>0$ is a small constant for numerical stability. These weights allow the fitting procedure to automatically trust low-error points more. In this study, we perform isotonic regression and let $\texttt{iso-reg}(\cdot)$ be the regressor model. The regressor can be regulated by a constraint, which enforces the monotonicity of the curve $R_{\mathrm{dyn}}(s)$ as a function of SoH $s$. To facilitate the regression, we sort the input points in the ascending order of SoH. Then, the regression process can be represented as,
\begin{equation}
\tilde{R}_{\mathrm{dyn}}(s)=\texttt{iso-reg}(\{(\hat{s}_i,R_{\mathrm{dyn}}^i),w_i\}_{i=1}^{c};\theta_{\mathrm{dyn}}),
\end{equation}
where $\theta_{\mathrm{dyn}}$ is the \texttt{Boolean} parameter of the monotonicity constraint. The regressor optimizes based on the weighted squared errors, and the model produces a non-decreasing fitted curve $\tilde{R}_{\mathrm{dyn}}(s)$ in the SoH-domain. 

The SoH-domain mapping for $R_{\mathrm{W}}(s)$ follows the same pipeline as $R_{\mathrm{dyn}}(s)$, except the weights calculation. Because $R_{\mathrm{W}}$ captures tail loss that is observed primarily near end-of-discharge, we assign a high (small) weight $w'_i$ to cycle $i$ in which the low-voltage tail occupies a large (small) fraction of the discharge duration. The tail and fraction can be determined by the same low-voltage gate during per-cycle LMS-based $R_{\mathrm{W}}$ identification. This allows the monotonic fit to trust $R_{\mathrm{W}}^{i}$ more when the tail effect is better supported by the data of cycle $i$. Here, the regression process for $R_{\mathrm{W}}$ is given by,
\begin{equation}
\tilde{R}_{\mathrm{W}}(s)=\texttt{iso-reg}(\{(\hat{s}_i,R_{\mathrm{W}}^i),w_i'\}_{i=1}^{c};\theta_{\mathrm{W}}),
\end{equation}
where $\theta_{\mathrm{W}}$ is the \texttt{Boolean} parameter of monotonicity.

\begin{figure*}
    \centering	
    \def \tmph{1.5in}
    \subfigure[$R_{\mathrm{dyn}}$ in SoH-domain]{\includegraphics[height=\tmph]{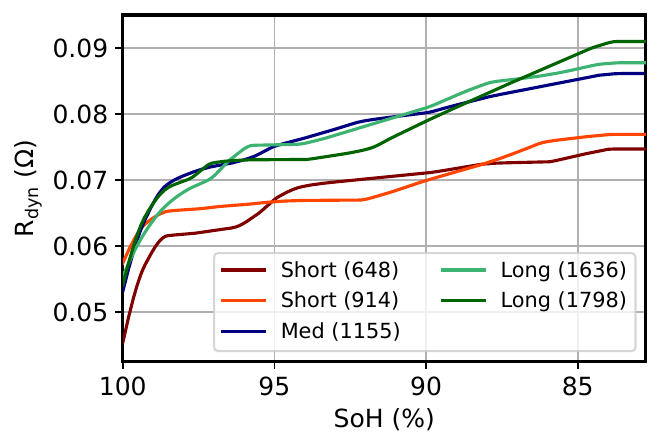}\label{fig:rdyn}}
    \hspace{-0em}
    \subfigure[$R_{\mathrm{W}}$ in SoH-domain]{\includegraphics[height=\tmph]{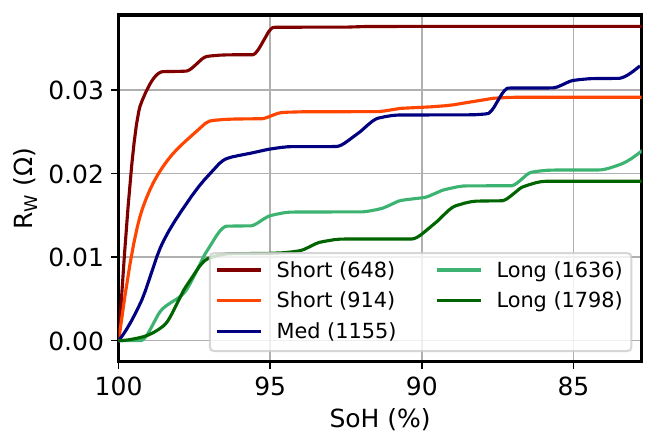}\label{fig:rw}}
    \hspace{-0em}
    \subfigure[$R_{\mathrm{dyn}}$ vs. $R_{\mathrm{W}}$]{\includegraphics[height=\tmph]{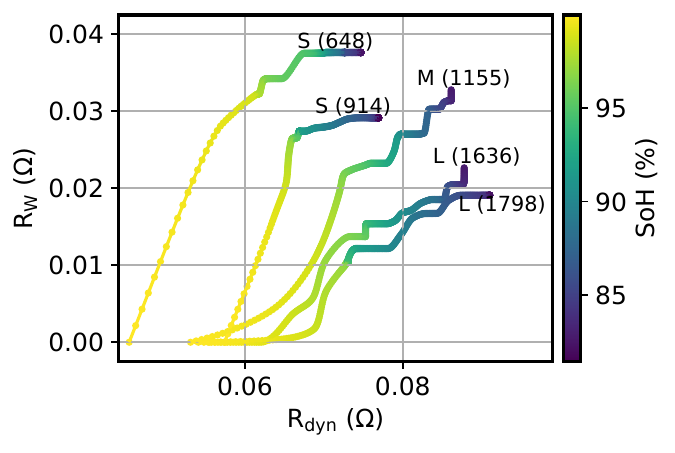}\label{fig:rdyn-w}}
    \caption{\textsc{IBAM} learned degradation fingerprints for representative batteries in the short-life (\texttt{Bat039} and \texttt{Bat025}), medium-life (\texttt{Bat116}), and long-life (\texttt{Bat100} and \texttt{Bat124}) categories; each battery’s lifespan (in cycles) is indicated in the legend. Overall, short-life batteries exhibit stronger tail loss.}
    \label{fig:r}
\end{figure*}

After fitting, we produce a compact representation that can be queried online via a lookup table. We uniformly sample the SoH axis over the valid operating range and configure $k$ reference SoH points, e.g., $k=25$ evenly spaced values over SoH from ideal condition 100\% to end-of-life (EoL) 80\%. During operation, given an inferred SoH $s$, e.g., by BiGRU, we locate the two neighboring reference points $s_i$ and $s_{i+1}$ and compute $R_p(s)$ by standard linear interpolation between $R_p(s_i)$ and $R_p(s_{i+1})$ in the lookup table for $p\in \{\mathrm{dyn},\mathrm{W}\}$.

\section{Experimental Study}
\label{sec:exp}

\subsection{Experiment Setup}
In our experiments, we utilize the MIT-Stanford battery dataset \cite{severson2019data} which is commonly used in battery studies. The dataset contains commercial cylindrical lithium-ion batteries with nominal capacity 1.1 Ah and voltage 3.3 V. The batteries were tested in a 30°C chamber via a cycler and discharged at a uniform 4C rate. For each battery and cycle, we focus on the battery operation scenarios and the discharge segment only. We standardize the start time of discharge as 0 and resample the entire discharge duration with 150 evenly distributed timestamps for consistency. We create three categories for representative batteries with short-, medium-, and long-lifespans. ML models are trained and evaluated using the processed discharge sequences, and optimized with the \texttt{AdamW} optimizer. \textsc{IBAM} is lightweight and experiments are implemented in Python 3.13 with PyTorch 2.7.1 and run on a CPU-only machine with Intel Core i5-220H and 24 GB RAM.

\subsection{Interpretable Fingerprint Insights}
In this part, we apply \textsc{IBAM} to extract degradation fingerprints of the tested batteries for health interpretation. For visualization clarity, we present representative batteries from each lifespan category. We select two short-life batteries, \texttt{Bat039} and \texttt{Bat025}, which reach EoL at cycles 648 and 941, respectively. The medium-life category is represented by \texttt{Bat116} with 1,155 cycles. The long-life category is represented by \texttt{Bat100} and \texttt{Bat124}, which reach EoL at cycles 1,636 and 1,798, respectively. We visualize the learned fingerprints by plotting $R_{\mathrm{dyn}}$ versus SoH, $R_{\mathrm{W}}$ versus SoH, and $R_{\mathrm{W}}$ versus $R_{\mathrm{dyn}}$ in Figs. \ref{fig:rdyn}, \ref{fig:rw}, and \ref{fig:rdyn-w}, respectively. These visualizations allow us to compare the trends of degradation mechanisms and we have the following observations.

First, we observe that both fingerprint components $R_{\mathrm{dyn}}$ and $R_{\mathrm{W}}$ increase as the battery ages. As shown in Fig. \ref{fig:rdyn}, $R_{\mathrm{dyn}}$ rises as SoH decreases, which indicates a growing polarization loss. Similarly, Fig. \ref{fig:rw} shows that $R_{\mathrm{W}}$ increases with aging, indicating an increasingly severe tail loss during the end-of-discharge. Importantly, both mechanisms can be observed throughout the tested health range. Even at high SoH, the batteries already exhibit measurable polarization loss and a non-zero tail loss. The fingerprint trends suggest that battery aging is not a single-mode process and different losses intensity concurrently. Among the two losses, $R_{\mathrm{W}}$ exhibits more significant lifespan discriminability than $R_{\mathrm{dyn}}$, which suggest that $R_{\mathrm{W}}$ can be a key differentiator of early EoL.

\begin{table*}[t]
\centering
\caption{The performance of different physics-based battery models, including ECM, FOECM$_{\text{base}}$ without the tail component $R_{\mathrm{W}}$, and \textsc{IBAM}'s FOECM in terms of per-cycle voltage-fitting RMSE. Four representative battery operation stages are considered, including 100\%, 67\%, 33\%, and 0\% remaining lifespan based on 100\% SoH and EoL. A small value indicates a good performance.}
\label{tab:comp}
\renewcommand{\arraystretch}{1.1}
\begin{tabular}{c|ccc|ccc|ccc|ccc} \hline \hline
\multirow{2}{*}{\shortstack{Optimized Physics\\Battery Model}} & \multicolumn{3}{c|}{100\% Lifespan} & \multicolumn{3}{c|}{67\% Lifespan} & \multicolumn{3}{c|}{33\% Lifespan} & \multicolumn{3}{c}{0\% Lifespan}  \\ \cline{2-13}
 & Short & Medium & Long & Short & Medium & Long & Short & Medium & Long & Short & Medium & Long \\ \hline
ECM & 53.067 & 48.350 & 46.723 & 39.073 & 34.607 & 29.777 & 42.073 & 37.133 & 32.140 & 39.813 & 38.577 & 32.663 \\ \hline
FOECM$_{\text{base}}$ (w/t $R_{\mathrm{W}})$ & 52.740 & 48.053 & 46.053 & 37.997 & 33.890 & 28.350 & 41.167 & 36.340 & 30.763 & 38.627 & 37.850 & 31.293 \\ \hline
FOECM (\textsc{IBAM}, ours) & \textbf{45.457} & \textbf{43.547} & \textbf{43.303} & \textbf{34.790} & \textbf{31.340} & \textbf{27.400} & \textbf{36.577} & \textbf{33.387} & \textbf{28.310} & \textbf{34.050} & \textbf{34.200} & \textbf{28.727} \\ \hline
\hline
\end{tabular}%
\end{table*}

Our next observation is a consistent fingerprint trade-off that aligns with lifespan. Long-life batteries are polarization dominant and relatively tail-healthy, whereas short-life batteries suffer more from tail loss. At a comparable health level, e.g., 90\% SoH, the two short-life batteries \texttt{Bat039} and \texttt{Bat025} exhibit low $R_{\mathrm{dyn}}$ of 0.071 and 0.070 $\Omega$, respectively, and 0.038 and 0.028 $\Omega$, respectively, for $R_{\mathrm{W}}$. In comparison, the long-life batteries \texttt{Bat100} and \texttt{Bat124} show higher $R_{\mathrm{dyn}}$, e.g., 0.081 and 0.079 $\Omega$, yet substantially lower $R_{\mathrm{W}}$, e.g., 0.017 and 0.012 $\Omega$ at the same SoH. The trajectories of $R_{\mathrm{dyn}}$ and $R_{\mathrm{W}}$ of the medium-life battery \texttt{Bat116} generally lie between the trajectories of the short- and long-life batteries. These results suggest that, at similar health levels, long-life batteries primarily incur a broad and curve-wide polarization penalty for discharge voltage while maintaining a strong end-of-discharge performance. Short-life batteries develop a pronounced tail bottleneck that can limit practical energy delivery even when SoH remains comparable. Importantly, a tail bottleneck has a threshold-like operational impact. As tail loss grows, the voltage near the end-of-discharge drops faster under the same load, so the battery reaches the cutoff voltage earlier and discharge terminates sooner. This means that usable battery energy is increasingly constrained by early cutoff. In contrast, long-life batteries' low $R_{\mathrm{W}}$ suggests that they sustain the end-of-discharge voltage for longer, so the batteries' degradation shows more as distributed loss rather than abrupt tail losses. 

Our third observation is that the proposed 2-D fingerprint offers an informative and intuitive visualization of degradation mechanisms, as shown in Fig. \ref{fig:rdyn-w}. Compared with tracking $R_{\mathrm{dyn}}(s)$ and $R_{\mathrm{W}}(s)$ separately along the SoH axis, the joint view in the $(R_{\mathrm{dyn}},R_{\mathrm{W}})$ plane provides a compact health portrait that distinguishes batteries with different lifespans more clearly. In particular, long-life batteries concentrate in a region with relatively high $R_{\mathrm{dyn}}$ but low $R_{\mathrm{W}}$, whereas short-life batteries extend toward much larger $R_{\mathrm{W}}$. Moreover, the SoH-colored trajectories reveal distinct aging trajectories. Short-life batteries follow a relatively vertical path  with rapid growth in $R_{\mathrm{W}}$, while long-life batteries evolve more uniformly. Overall, the 2-D fingerprints enable an efficient comparison of degradation mechanisms beyond a single SoH.

A practical implication of the 2-D fingerprints is that they suggest different battery operation priorities even at the same level of battery health. For batteries dominated by tail loss, a main risk is early cutoff near end-of-discharge. Thus, it suggests conservative battery operations when the battery capacity is low, e.g., avoid deep discharge and apply a higher practical cutoff voltage under load. In contrast, different operation aspects shall be considered for polarization-dominant batteries, e.g., limiting peak current through non-fast charging/discharging or current derating to minimize the associated battery stress. Overall, \textsc{IBAM} provides a mechanism-aware basis with interpretability for optimizing battery operations.

\subsection{Physics-Model Fitting Performance}
In this part, we evaluate the performance of different physics-based battery models. We identify and optimize their key parameters with \textsc{IBAM} and the models' performance is reflected by comparing the models' per-cycle output voltage and the measured voltage from routine BMS logs. Specifically, we consider a ECM with one resistor–capacitor pair, FOECM$_{\text{base}}$ without the the Warburg diffusion element and $R_{\mathrm{W}}$, and a full-version FOECM with both $R_{\mathrm{dyn}}$ and $R_{\mathrm{W}}$. We report the models' voltage-fitting error in terms of root mean square error (RMSE), evaluated at four lifespan stages, i.e., 100\%, 67\%, 33\%, and 0\% of the lifespan, the first and the last of which denote the cycle with 100\% SoH and the EoL cycle, respectively. The results are presented in Table \ref{tab:comp}.

First, \textsc{IBAM}'s optimized FOECM achieves the best fidelity. As shown in the table, FOECM yields the lowest fitting error in all reported settings covering four lifespan stages compared to ECM. For the short-lifespan batteries at their EoL cycles, ECM has an RMSE of 39.8 and the RMSE of FOECM is 34.1, which means a 14.5\% error reduction. When we consider the best health with 100\% SoH for the same battery category, FOECM can reduce the error by 14.3\%. In addition, FOECM$_{\text{base}}$, i.e., without capturing tail loss $R_{\mathrm{W}}$, can reduce the errors of ECM, but still shows significant performance gaps compared to FOECM. For the two settings above, the errors of FOECM$_{\text{base}}$ are 38.6 and 52.7, respectively, with 3.0\% and 0.7\% error reductions over ECM, which are much lower than the percentage reductions achieved by FOECM. The results show that simply switching from ECM to a basic FOECM offers limited gains, and the full \textsc{IBAM} design contributes the dominant improvement in modelling fidelity. 

We also observe that \textsc{IBAM}'s performance advantages are more evident when batteries are close to their EoL. This is visible in the table when \textsc{IBAM} FOECM is compared against ECM and FOECM$_{\text{base}}$ in different lifespan stages. For short-life batteries at 67\% remaining lifespan, \textsc{IBAM} improves over ECM and FOECM$_{\text{base}}$ with 11.0\% and 8.4\% error reductions, respectively. At the batteries' EoL cycles, the reductions enlarge to 14.5\% and 11.8\%, respectively. Similar patterns can be observed for medium- and long-life batteries, e.g., the percentages increases from 8.0\% and 3.4\% at 67\% remaining lifespan to 12.1\% and 8.2\% at the EoL cycles. The results show that the benefit of \textsc{IBAM} FOECM is not a constant, and the solution becomes more valuable when batteries are more aged, where voltage behavior can be harder to fit. Meanwhile, FOECM$_{\text{base}}$ performs relatively similar to ECM, which implies that the late-life performance gain of \textsc{IBAM} does not come from just switching ECM to FOECM. As the voltage behavior distorts near EoL, \textsc{IBAM}'s explicit modelling of the tail terms becomes increasingly important.

Besides, voltage fitting is more challenging for short-life batteries than long-life ones and the errors of short-life batteries are higher than the rest in most of the reported settings. For example, ECM reduces the errors from 53.1 for short-life batteries to 46.7 for long-life batteries when the batteries are new. For aged batteries, ECM can still reduce the errors from 39.8 to 32.7. \textsc{IBAM}'s FOECM achieves 43.3 RMSE for long-life batteries, lower than 45.5 for short-life batteries at 100\% lifespan. When one-third of the lifespan is left, \textsc{IBAM}'s error is 28.7 for long-life batteries, smaller than 34.1 for short-life batteries. The results imply that short-life batteries exhibit voltage dynamics that is more difficult to explain using the same model and parameter identification process. This is consistent with our previous observations about short-life batteries' increased tail effect and potentially strong variability. On the contrary, long-life batteries relatively follow regular degradation patterns that are easier to be modeled.

\begin{table}[t]
\centering
\caption{Impact of 4-channel feature construction on the BiGRU SoH estimator used in \textsc{IBAM}. Results are reported in MAE and RMSE, where small values indicate good performance.}
\label{tab:gru}
\renewcommand{\arraystretch}{1.1}
\begin{tabular}{c|cc|cc} \hline \hline
\multirow{2}{*}{\shortstack{BiGRU Input\\Battery Lifespan}} & \multicolumn{2}{c|}{MAE} & \multicolumn{2}{c}{RMSE}  \\ \cline{2-5}
 & 1-Channel & 4-Channel & 1-Channel & 4-Channel \\ \hline
Short & 0.0141 & \textbf{0.0039} & 0.0174 & \textbf{0.0050}  \\ \hline
Medium & 0.0184 & \textbf{0.0053} & 0.0216 & 
\textbf{0.0060} \\ \hline
Long & 0.0111 & \textbf{0.0056} & 0.0132 & \textbf{0.0066}  \\ \hline
\hline
\end{tabular}%
\end{table}

\subsection{Impact of Multi-Channel Inputs on BiGRU SoH Modelling}
In \textsc{IBAM}, SoH is not only a prediction target but also the health coordinate used to construct and query the SoH-domain fingerprints. Therefore, errors in SoH estimation propagate directly into the fingerprints. This can blur the SoH-fingerprint relations across cycles and reduce the quality of interpretation. As such, we conduct an ablation study on the BiGRU-based SoH estimator, specifically on the impact of our designed 4-channel feature construction to justify the design for \textsc{IBAM}. Here, we keep the BiGRU backbone and training protocol fixed and vary the input representation. The baseline model takes the raw discharge voltage sequence of each cycle as a 1-channel feature as input. In \textsc{IBAM}, we construct a 4-channel input with the raw voltage, relative voltage, voltage change rate, and timestamps. We evaluate SoH modelling performance using mean absolute error (MAE) and RMSE and report the results in Table \ref{tab:gru} for batteries in different lifespan categories.

First, we can see that the proposed 4-channel feature construction consistently improves BiGRU SoH modelling accuracy over using a single raw-voltage channel. For example, MAE drops from 0.014 with 1-channel input to 0.004 with 4-channel input for short-life batteries, where RMSE decreases from 0.017 to 0.005. For medium-life batteries, 4-channel input helps reduce the errors by 71.2\% and 72.2\% for MAE and RMSE, respectively. The improvements are expected because the multi-channel features expose complementary battery health patterns that a single voltage trace is not able to provide. For example, the absolute voltage reflects the overall discharge profile and the voltage change rate captures local dynamics of battery operation. Together, multiple channels provide a more informative representation and allow the lightweight BiGRU to learn health-related patterns effectively. The accurate SoH estimates directly benefit \textsc{IBAM}, by serving as accurate coordinates of the health fingerprints.
%

\balance
We also observe a mild trend that SoH modelling becomes more challenging for batteries with longer lifespans. \textsc{IBAM}'s SoH modelling MAE increases from 0.004 for short-life batteries to 0.006 for long-life batteries, and RMSE increases from 0.005 to 0.007. A possible reason is that long-life batteries tend to exhibit slow degradation, which brings challenges to distinguish nearby SoH levels between per-cycle voltage features. Besides, long degradation trajectories may accumulate battery operation variability over time, which increases diversity of the long-life category and thus modelling errors even with the same ML backbone. Nevertheless, the errors remain low in \textsc{IBAM} and are consistently small, suggesting that \textsc{IBAM} remains robust even for more challenging SoH modelling. 

\section{Conclusion}
\label{sec:conclusion}
In this paper, we propose \textsc{IBAM} for interpretable battery aging modelling from routine BMS logs. \textsc{IBAM} constructs a physics-based battery model and extracts 2-D aging fingerprints from the model to quantify polarization and tail losses. To obtain the fingerprints accurately, \textsc{IBAM} performs per-cycle parameter fitting with a two-stage LSM. It then estimates cycle-wise SoH via BiGRU with multi-channel voltage features, and produces physically consistent SoH-indexed fingerprints. Experiments on batteries with different lifespans show that \textsc{IBAM} consistently improves modelling fidelity and offers interpretable fingerprint patterns that are not visiable from SoH alone. Overall, \textsc{IBAM} bridges accurate battery health estimation and actionable interpretability by turning routine BMS logs into degradation mechanism indicators, and can improve BMS operations and enable safe battery usage.

\bibliographystyle{IEEEtran}
\bibliography{reference}

\begin{thebibliography}{10}
\providecommand{\url}[1]{#1}
\csname url@samestyle\endcsname
\providecommand{\newblock}{\relax}
\providecommand{\bibinfo}[2]{#2}
\providecommand{\BIBentrySTDinterwordspacing}{\spaceskip=0pt\relax}
\providecommand{\BIBentryALTinterwordstretchfactor}{4}
\providecommand{\BIBentryALTinterwordspacing}{\spaceskip=\fontdimen2\font plus
\BIBentryALTinterwordstretchfactor\fontdimen3\font minus \fontdimen4\font\relax}
\providecommand{\BIBforeignlanguage}[2]{{%
\expandafter\ifx\csname l@#1\endcsname\relax
\typeout{** WARNING: IEEEtran.bst: No hyphenation pattern has been}%
\typeout{** loaded for the language `#1'. Using the pattern for}%
\typeout{** the default language instead.}%
\else
\language=\csname l@#1\endcsname
\fi
#2}}
\providecommand{\BIBdecl}{\relax}
\BIBdecl

\bibitem{IEA25}
{IEA}, ``Iea global ev outlook 2025,'' \url{https://www.iea.org/reports/global-ev-outlook-2025}, 2025, accessed: 2026-01-29.

\bibitem{severson2019data}
K.~A. Severson, P.~M. Attia, N.~Jin \emph{et~al.}, ``Data-driven prediction of battery cycle life before capacity degradation,'' \emph{Nature Energy}, vol.~4, no.~5, pp. 383--391, 2019.

\bibitem{kokkalis2024predicting}
K.~Kokkalis, C.~Chronis, E.~Politi \emph{et~al.}, ``Predicting the lifespan of lithium-ion batteries using machine learning, parameter tuning and model ensembles,'' in \emph{2024 IEEE 10th International Conference on Big Data Computing Service and Machine Learning Applications (BigDataService)}.\hskip 1em plus 0.5em minus 0.4em\relax IEEE, 2024, pp. 122--129.

\bibitem{liu2025data}
Z.~Liu, Y.~Liu, Y.~Zhang \emph{et~al.}, ``Data-driven lithium-ion battery soh prediction: A novel shmm-transformer-bigru hybrid neural network method,'' \emph{Measurement}, p. 118579, 2025.

\bibitem{sameer2025ginet}
S.~Sameer, W.~Zhang, X.~Lou \emph{et~al.}, ``Ginet: Integrating sequential and context-aware learning for battery capacity prediction,'' \emph{arXiv preprint arXiv:2501.04997}, 2025.

\bibitem{li2026entrolnn}
W.~Li, W.~Zhang, and Q.~Yan, ``Entrolnn: Entropy-guided liquid neural networks for operando refinement of battery capacity fade trajectories,'' \emph{arXiv preprint arXiv:2601.06195}, 2026.

\bibitem{sameer2025pace}
S.~Sameer, W.~Zhang, K.~D. Dharshini \emph{et~al.}, ``Pace: Physics-aware attentive temporal convolutional network for battery health estimation,'' \emph{arXiv preprint arXiv:2512.11332}, 2025.

\bibitem{eivazi2024diffbatt}
H.~Eivazi, A.~Hebenbrock, R.~Ginster \emph{et~al.}, ``Diffbatt: A diffusion model for battery degradation prediction and synthesis,'' \emph{arXiv preprint arXiv:2410.23893}, 2024.

\bibitem{kwan5871665knowdiff}
Z.~Kwan, W.~Zhang, A.~B. Ng \emph{et~al.}, ``Knowdiff: Knowledge-regulated dual-stream diffusion for robust battery prognostics,'' \emph{Available at SSRN 5871665}.

\bibitem{lim2025unified}
E.~I. H.-E. Lim and N.~H.~L. Wong, ``A unified framework for interpretable and uncertainty-aware battery state of health estimation using deep neural networks,'' in \emph{2025 Asia Pacific Signal and Information Processing Association Annual Summit and Conference (APSIPA ASC)}.\hskip 1em plus 0.5em minus 0.4em\relax IEEE, 2025, pp. 1392--1397.

\bibitem{gahramanov2025detecting}
O.~Gahramanov, H.~Hokmabad, E.~Ginzburg-Ganz \emph{et~al.}, ``Detecting battery degradation factors using explainable ai,'' in \emph{2025 IEEE PES Innovative Smart Grid Technologies Conference Europe (ISGT Europe)}.\hskip 1em plus 0.5em minus 0.4em\relax IEEE, 2025, pp. 1--5.

\bibitem{tetik2026feature}
T.~Tetik, ``Feature engineering and explainable artificial intelligence for state of health estimation of lithium-ion batteries,'' \emph{Journal of Energy Storage}, vol. 144, p. 119873, 2026.

\bibitem{liu2024ultrasonic}
K.~Liu, Y.~Liu, S.~Zhao \emph{et~al.}, ``An ultrasonic wave-based method for efficient state-of-health estimation of li-ion batteries,'' \emph{IEEE Transactions on Industrial Electronics}, 2024.

\bibitem{xie2025impedance}
Y.~Xie, W.~Guo, T.~Zhou \emph{et~al.}, ``An impedance-based electro-thermal model integrated with in-situ lithium-plating criterion for ac heating at low temperatures,'' \emph{Applied Energy}, vol. 391, p. 125930, 2025.

\end{thebibliography}

\end{document}